# Parallelizing Optimal Multiple Sequence Alignment by Dynamic Programming

Manal Helal[1], Hossam El-Gindy[1], Lenore Mullin[2], Bruno Gaeta[1]

[1] School of Computer Science and Engineering, Faculty of Engineering, University of New South Wales, Sydney, Australia
[2] National Science Foundation, Washington, USA
mhelal@cse.unsw.edu.au

## Abstract

*Optimal multiple sequence alignment by dynamic programming, like many highly dimensional scientific computing problems, has failed to benefit from the improvements in computing performance brought about by multi-processor systems, due to the lack of suitable scheme to manage partitioning and dependencies. A scheme for parallel implementation of the dynamic programming multiple sequence alignment is presented, based on a peer to peer design and a multidimensional array indexing method. This design results in up to 5-fold improvement compared to a previously described master/slave design, and scales favourably with the number of processors used. This study demonstrates an approach for parallelising multi-dimensional dynamic programming and similar algorithms utilizing multi-processor architectures.*

## 1. Introduction

Parallel computation has been limited to high performance computing environments. Now it is becoming widespread even in consumer-level computers, while an increasing number of homogeneous, heterogeneous and hybrid architectures harnessing the power of multiple processors are being developed. However, many compute-intensive problems have not benefited from multiple processor architectures even when the algorithm is parallelisable, due to difficulties in managing dependencies and memory indexing. Existing compilers RARELY extract such parallelism adequately.

We investigate the use of multidimensional array indexing methods. Such methods allow us capitalize on data structures, their algebra and layout, and access patterns up and down the processor/memory hierarchy. This is possible given the contiguity of tensors, e.g. n-d array layouts. Our objective is to provide a unified partitioning scheme and dependency modelling with inherent load balancing that delivers solutions for algorithms like high dimensional dynamic programming (DP). The present work provides a framework for harnessing complex processor/memory hierarchies in computer architectures. We are presenting a distributed Peer-to-Peer (P2P) design for the DP algorithm for the multiple sequence alignment (MSA) problem in computational biology that performs up to five times better than a previous Master/Slave design and scores better than clustalW.

## 2. Dynamic Programming algorithm for Multiple Sequence Alignment

Multiple Sequence Alignment (MSA) is an example of a highly dimensional problem whose optimal solution is difficult to parallelize. The computational complexity of the optimal DP-MSA algorithm has led to the development of a number of heuristic solutions that are widely used in biology and genomics research. The study in [1] shows that when applied repeatedly on thousands of genes, the alignment can lead to several problems, including different alignment methods resulting in different conclusions. The assessment in [2] shows that using different MSA programs is advisable to eliminate bias and incorrect gap placement in the alignments.

The DP algorithms are proven in [3] to be mathematically optimal, efficient and indiscriminate. Sequence alignment by dynamic programming is commonly used for aligning two sequences. However, in practice DP is difficult to apply to multiple sequence alignment because the scoring tensor grows exponentially with the number of sequences in the dataset.

It is prohibitive to use DP algorithms for highly dimensional search spaces on a single machine, and / or with full search space. The increase in space and time requirements cannot be solved by recruiting additional processors due to the lack of indexing methods that can address the scoring tensor invariant of shape and dimension. When aligning two sequences by dynamic programming, the sequences are stretched on the two axes of a scoring matrix. All possible pairs of characters are matched by following a scoring scheme for matches, mismatches and gaps scores. This process generates a matrix of numbers that represent all possible alignments. The optimal alignment can be found by tracing back, starting from the highest score on the bottom edges of the matrix, and following the highest scores on the matrix up to the origin.

The pair-wise sequence alignment dynamic programming algorithm uses the following recurrence to score the matrix:

**Equation 1: Dynamic Programming 2D scoring recurrence**

$$\xi_{i,j} = \max \begin{pmatrix} \xi_{i-1,j-1} + sub(a_i, b_j) \\ \xi_{i-1,j} + g \\ \xi_{i,j-1} + g \end{pmatrix}$$

where $\xi$ is the scoring matrix, i and j are the vertical and horizontal indices, a and b are the two input sequences, sub is the substitution score of 2 residues read at positions i and j from sequences a and b respectively, based on the used scoring scheme.

For two sequences, three neighbours need to be checked for scoring each cell. For k sequences (k-dimensional MSA) we need to check $2^k-1$ neighbours for computing each cell score in a tensor of k dimensionality containing $\tau\rho\xi$ cells (the product "$\tau$" of the shape "$\rho$" of the tensor (lengths of input sequences)). This leads to $O((2^k-1) * (\tau\rho\xi))$ computation complexity, where the search space is exponentially growing with the input data size.

This work investigates the design of an optimal DP-MSA employing parallel programming to distribute the k-dimensional scoring tensor over multiple processors. The choice of DP algorithm was made to address the bias and low quality of alignments.

This approach requires a multidimensional index representation invariant of shape and dimension, a suitable and efficient partitioning/scheduling method and management of dependencies. We adopted a representation based on Mathematics of Arrays (MoA) as the basis for our indexing scheme.

## 3. Optimal MoA based DP-MSA Peer to Peer (P2P) Solution

We have previously presented a master/slave design for an optimal distributed DP-MSA algorithm [5] based on a master process responsible for dividing the DP-MSA k-dimensional scoring tensor into partitions that are distributed to individual processors. The dependencies and communication between partitions were managed using MoA, a formalism offering a set of constructs to represent and manipulate highly multidimensional arrays in memory in a linear, concise and efficient way, invariant of data dimensionality and shape [4]. This is achieved through an algebraic restructuring of the existing expressions that adapt to the processor/memory hierarchies. The partitions needed to be reasonably sized so that they can be scored simultaneously in parallel by as many processors as possible, while keeping the communication cost low. This concept is applicable in shared memory, distributed memory and hybrid shared/distributed memory processor memory hierarchies. A generic granularity design was proposed, where communication and computation costs can be optimized to minimize the total execution time. The granularity of the design is decided by optimizing the partitioning size S, to the size of the data set read, and the capabilities of the HPC machine used (computation cost vs. communication cost). Thus, different levels, e.g. processors, nodes, …, contribute to speed, size, time calculations and can be calculated based on access patterns.

Different scheduling schemes were tested and dependency-based scheduling (closest indexed partitions clustered together in one processor) was used to reduce the communication. Experimental evaluation of the model revealed the characteristics of the true parallelism of the problem. According to the computation wave definition in [8], it was found that not all $(2^k - 1)$ needed lower indexed neighbours are computed in one wave of computation and then passed in the next wave to the higher neighbouring cell to be computed. The process blocks until the remote lower neighbouring cell scores are received on k waves of computations at most, because the neighbours as retrieved by the MoA neighbours equation defined in [8] belong to different partitions at different distances from the origin. A partition is defined by the global index of its first cell, referred to as partition index. The distance from the origin is defined as the sum of the partition index vector elements divided by (S-1) where S is the partition size, compared to the origin cell. Hence, the partitions that can be simultaneously scored, and do not require dependency from each other in the same computation wave, are divided into groups of equal distance from the origin, to be computed in one wave. Then dependencies are communicated, and partitions at a higher distance can be computed, having received the previous dependencies from the previous wave.

This study redefines the computation wave definition to be described as all partitions located at the same distance from the origin that can be executed simultaneously by the available processors. This term indicates the grouping of independent partitions, and the motion to the next wave. MPI clock-wise (higher ranked processors send to lower ranked processors) non-blocking communication is used, then a barrier to synchronize the communication. Then an anti-clockwise communication (lower ranked processors send to higher ranked processors) takes place. This guarantees all processors are computing the same computation wave then communicate all required dependencies for the next wave. This method also eliminates deadlocks, as no two processors would be involved in two-way communication at the same time.

The P2P term used here refers to the partitioning method. The partitioning is pre-processed, and each processor can fetch its own partitions, and calculate the destination processor ID to send the dependency to. There is no need for a scheduler master process to partition and manage dependency communication, eliminating the extra master's computation and communication cost.

### 3.1 P2P Partitioning Formalization

For a given data set of k sequences ($\delta\xi$ - dimensionality), and of shape $\rho$: (sequences lengths) $\rho_i$ length ($0 \leq i < k$), and given a partitioning size S, the total number of waves t for a given dataset (which is the maximum distance from the origin as multiples of the partition size chosen), is calculated by equation 2.

**Equation 2: Total Number of waves**

$$t = \frac{\rho_0 - 1}{S - 1} + \sum_{i=1}^{k-1}\left(\frac{\rho_i - 1}{S - 1} - 1\right)$$

The total number of partitions P all over the waves based on a partition size S, is the product of total number of partitions created in each dimension, and can be calculated in equation 3:

**Equation 3: Total Number of Partitions**

$$P = \prod_{i=0}^{k-1} p_i$$

Where $p_i = \frac{\rho_i - 1}{S - 1}$

The number of overlapping cells C between partitions that will be computed in one partition and sent to other partitions for dependency, are calculated in equation 4.

**Equation 4: Communication cost - dependency equation**

$$C = \sum_{i=0}^{k-1} C_i$$

$$C_0 = p_0 - 1$$

Where

$$C_i = C_{i-1} \times \rho_i + \left(\prod_{j=0}^{i} p_j\right) \times 2^i - 1$$

$w_i$ is the $i^{th}$ wave where $0 \leq i < t$. In each $w_i$ there will be $p_i$ partitions, distributed equally between the V available processors. The scheduling (processor assignments) is a simple rounded division as shown in equation 5.

**Equation 5: Scheduling Equation**

$$m_{p_{j,i}} = \left\lceil \frac{p_i}{V} \right\rceil$$

Where $0 \leq m < V$, and $m_{pj,i}$ is the processor that scores the $j^{th}$ partition in the $i^{th}$ wave.

### 3.2 Wave Calculation

Waves and their partitions are pre-processed and each processor fetches its partitions in the current computation wave $w_i$, based on the partition order, $p_{ij}$ the total number of partitions in the current computation wave: pi, and the number of processors in the cluster V, as formalized in equation 5. The following algorithm assigns partitions indices to the ($p_{ij}$) 2D array.

```
getwavesPartitions (t, p)
w = 0
do
   do
      p[i][j] ← getNextIntegerPartition(w, more_IP)
      AddPartIndex (p[i][j])
      do
         p[i][j] ← permute (p[i][j], more_Perm)
         AddPartIndex (p[i][j])
      while more_Perm
   while more_IP
   i ← i + 1
while i < t
```

The algorithm loops for all waves *i* until the maximum wave *t* is reached. For each wave (distance from the origin) the integer partitions of the wave number (the different combinations of integers that will add up to the wave number) are retrieved. For examples: wave 2 has (1, 1) and (2, 0) for 2D index; wave 3 has (2, 1), and (3, 0) for 2D index, and (1, 1, 1), (2, 1, 0) and (3, 0, 0) for 3D index. Then the permutations of these indices are iterated. The function AddPartitionIndex, multiplies the index with the partition size, and check for the shape constraints (the lengths of the sequences – some integer partitions or permutations will not be valid if one index exceeds the maximum), then add the index to the $p_{ij}$ and increment j. Table 1 provides number of independent partitions that can be processed in parallel, preferably in different processors for maximum speed-up. Table 2 provide the actual partitioning of 4 sequences with lengths (8, 8, 8, 8), which will be (9, 9, 9, 9) after including the initial gap character, resulting in 6561 cells to be computed. Dividing by a partition size S = 3 in each dimension, we get 256 partitions in 13 waves according to the above equations. Figure 1 plots data in table 1, showing on the x-axis with logarithmic scale the number of partitions that can be scored in parallel over the waves in the y-axis, for 10 dimensions as shown in the legend.

**Table 1: Partition Parallelism (Total Partitions per wave) shown for 9 dimensionality in rows(k), and 9 waves in Columns.**

| k | Waves | | | | | | | | |
|---|---|---|---|---|---|---|---|---|---|
|   | 1 | 2 | 3 | 4 | 5 | 6 | 7 | 8 | 9 |
| 2 | 1 | 2 | 3 | 4 | 5 | 6 | 7 | 8 | 9 |
| 3 | 1 | 3 | 6 | 10 | 15 | 21 | 28 | 36 | 45 |
| 4 | 1 | 4 | 10 | 20 | 35 | 56 | 84 | 120 | 165 |
| 5 | 1 | 5 | 15 | 35 | 70 | 126 | 210 | 330 | 495 |
| 6 | 1 | 6 | 21 | 56 | 126 | 252 | 462 | 792 | 1287 |
| 7 | 1 | 7 | 28 | 84 | 210 | 462 | 924 | 1716 | 3003 |
| 8 | 1 | 8 | 36 | 120 | 330 | 792 | 1716 | 3432 | 6435 |
| 9 | 1 | 9 | 45 | 165 | 495 | 1287 | 3003 | 6435 | 12870 |

**Table 2: 4Dmof shape (9, 9, 9, 9) partitions ($p_w$) per wave (w).**

| w | 1 | 2 | 3 | 4 | 5 | 6 | 7 | 8 | 9 | 10 | 11 | 12 | 13 |
|---|---|---|---|---|---|---|---|---|---|----|----|----|----|
| $p_w$ | 1 | 4 | 10 | 20 | 31 | 40 | 44 | 40 | 31 | 20 | 10 | 4 | 1 |

**Figure 1: Partitions parallelism per wave for 10 dimensions.**

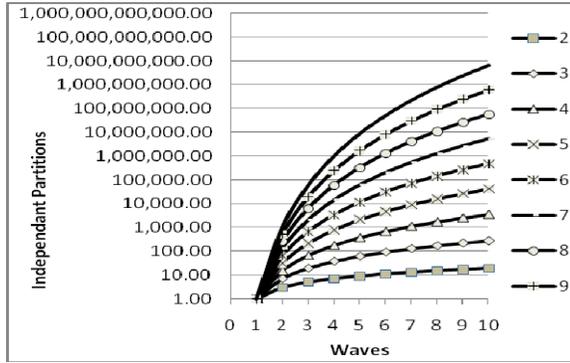

### 5.3 P2P Scoring and Dependency Communication

The same scoring scheme used in the master/slave design, and detailed in the complexity analysis section 6 is used in the P2P design. However, dependency is packed to be communicated in between waves of computation, and done by each processor independently, and not by a master process. In the current implementation, the dependency relationships are computed by retrieving the higher indexed neighbours of the current cell, their partition index and the corresponding processor. If the neighbour partition is found in a different processor, it is added to the OCout (Overlapping Cells outgoing) buffer, to be sent after all the cells in all partitions in the current computation wave are scored. Then all OCout cells are packed to be sent in one buffered MPI communication per processor to reduce the TCP/IP overhead. Figure 2 shows the dependency network for a 2D example, and a 3d example. The nodes (circles) in one level in the network represent partitions in one computation wave. The colour of the node represents the processors computing the partition, showing the dependency-based scheduling to reduce communication.

**Figure 2: Dependency networks connecting partitions in one wave to their dependent partition(s) in the next wave for 2D (a) and 3D (b).**

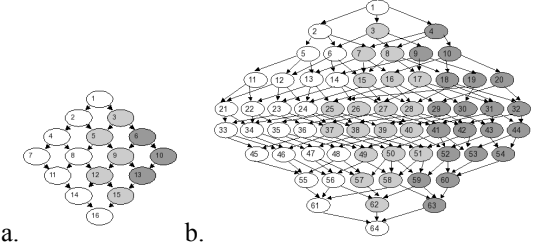

a.   b.

## 4. Complexity Analysis

The complexity of the proposed solution is analysed in two levels: the computation and the communication. Both are dependent on the partitioning size S, and load balancing. As mentioned above, the computation complexity of the scoring of the $l^{th}$ cell ($\xi_l$), where $0 \leq l < S^k$, in one partition ($p_{ij}$), in one computational wave ($w_i$), in one processor m, is as follows:

- Process the current cell based on its position:
  - If initial cell, initialize it ≡ O(1)
  - Otherwise:
    * If local cell (in another partition), check the (OCout) buffer for the previous wave ≡ $O(p_{(i-1),m} \times (S^k - (S-1)^k)) \equiv O(S^k)$
    * If not local: check received list (OCin) for the previous wave ≡ $O(p_{(i-1),m} \times (S^k - (S-1)^k)) \equiv O(S^k)$
    * Otherwise, block wait till received ≡ O(1)
  - If internal cell, compute the score as follows:
    * get pair-wise scores for each pair of residues ≡ $O(k^2)$
    * get lower neighbours' scores ≡ $O(2^k - 1)$
    * compute temporary Neighbour score ≡ $O(3 \times 2^k - 1)$
    * assign maximum temporary score to current cell score ≡ O(1)
- Dependency Analysis ≡ $O(p_{(i+1),m} \times (S^k - (S-1)^k)) \equiv O(S^k)$

The worst case complexity when scoring a cell up to the block wait step, is: $O(k)+O(S^k)+O(S^k)+O(1)+O(S^k) \equiv O(S^k)$, and the blocking time is dependent on the architecture.

After a cell is computed, a dependency analysis is conducted by retrieving the cell's higher indexed neighbours, and then retrieves each neighbour's partition index and the computing processor. Cell scores is added to the OCout buffer if the neighbour's processor is different from the current processor computing the cell. The execution time in a distributed environment is estimated by equation number (6) as defined by [6]:

**Equation 6: Distributed Execution Time Equation**

$$dT = r \times \max(p_m) + \left(\left(\frac{c}{2}\right) \times \left(P^2 \sum_{m=0}^{V-1} p_m^2\right)\right)$$

Where dT is the distributed overall time, composed of the maximum computation cost on a highest loaded processor plus the communication cost as a mesh of communication of each partition to each partition.

r is the execution time of a single partition on a processor, which is $O(S^k \times (2^k-1))$ for all cells in each partition, in each wave assigned to a processor m. It is multiplied by the maximum processor allocation $max(p_m)$. c is the communication cost between the single task and all other tasks on other processors, which is equivalent to $O(S^k – (S-1)^k)$ per partition. This assumes a mesh of communication between all partitions. However, figure 2 shows that each partition communicates with maximum $2^k-1$ other partitions that according to the scheduling scheme, most of them would be local. For the local partitions, the search cost could be equivalent. Equation 6 shows worst case communication. However, due to operating system scheduling of processors in a single multi-core HPC, or equivalently, network routing on a distributed cluster of computing nodes, it is hard to estimate a more accurate equation. P is the total partitions in all waves. The r and c are attributes of the hardware we are executing on. We can also calculate R, which is the total computation time for all partitions assigned to a processor and C which is total communication done by a processor for all partitions assigned to it. R and C can be easily calculated based on the partition size, and the ratio R/C is the granularity of the design. Minimizing the total execution time takes place by altering these ratios.

## 5. Results

### 5.1 P2P Design Performance

The P2P design was implemented in ANSI C using an MPI standard 2 library and compared to a similarly implemented Master/slave design on The APAC SGI Altix 3700 Bx2 (1928x1.6Ghz Itanium2 processors)[1]. The P2P design was tested on one processor (sequential) as well as on multiple processors (P2P). Evaluation results in terms of CPU time, elapsed time, physical and virtual memory use are shown in Figure 3, where the x-axis is the exponentially growing data size. The P2P implementation achieved up to five times speedup and better memory optimization as data size increased, relative to the master/slave design. Running the solution sequentially on one processor without any partitioning or communication and in full search space provided better performance than the parallel solution for small data sizes. However, when increasing the number and length of sequences being aligned, the P2P version achieved an almost 10-fold reduction in CPU time and 2-fold reduction in elapsed time on 4 processors, and 4-5 fold decrease in memory usage.

**Figure 3:** Performance evaluation of P2P parallel dynamic programming MSA relative to Master/Slave and sequential implementations showing the data size in the x-axis (product of sequences' lengths – cells to be scored).

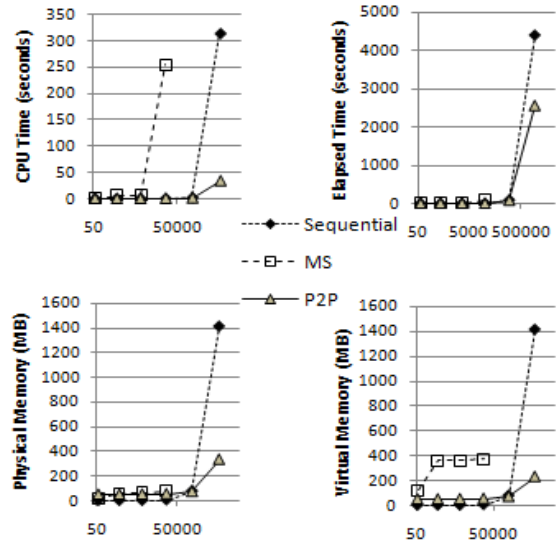

### 5.2 Comparison with heuristic methods

The dynamic programming alignment presented is proven mathematically to be optimal. Some tests were conducted to compare the results with existing heuristic MSA methods to validate the scaling of the scoring scheme used with dimensions. Running a 6 sequences test case in clustalW [7] using default parameters resulted in an alignment with similarity score = -36. Running the same test using the T-Coffee [8] online server resulted in an alignment with similarity score = -22. The MUSCLE [9] server on EBI website produced an alignment with similarity score = -12. The presented algorithm resulted in an alignment with a higher score of -5. Similarity scores were calculated using the sum of pairs method for each column, with +1 for a match, 0 for a mismatch, -1 for a gap and 0 for pairs of gaps. ClustalW, T-Coffee and Muscle are progressive methods and use position-specific gap penalties scoring methods. ClustalW and MUSCLE inserted the gaps all next to each other on one or both ends of the sequences. T-Coffee and the optimal algorithm distributed the gaps more evenly.

### 5.3 Processor Scalability

In order to measure processor scalability performance, the same datasets were used to test the performance on 4, 6, 8, 16, 24, 32 and 64 processors. The resulting elapsed time scalability profile is shown in Figure 4. The elapsed time decreases as the number of processors increases. This type of profile is very desirable for any distributed application.

[1] This work was supported by an award under the Merit Allocation Scheme on the National Facility of the Australian Partnership for Advanced Computing. (http://nf.apac.edu.au/facilities/ac/hardware.php)

**Figure 4: Processor Scalability**

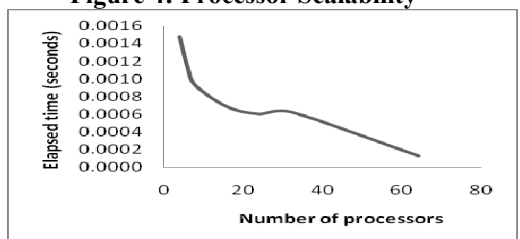

## 6. Conclusion

We developed a multi-processor P2P implementation of the MSA dynamic programming algorithm that resulted in marked performance improvements over sequential and master/slave implementations, and produced higher-scoring alignments than commonly used heuristic methods. The presented work contributes a unified reconfigurable partitioning scheme to divide computation complexity over processors, and model communication requirements. An automatic load balancing scheme was found to be inherent in the methods used by clustering partitions based on location. Each cluster of partitions is assigned to the same processor to reduce communication.

Parallelism scalability is growing exponentially with the data size, which allows for higher processor scalability. The upper bound of the data size is based on the configuration of the cluster, grid, or HPC used, and the number of processors used. We discussed our implementation using MPI knowing that this environment is supported in both shared and distributed memory platforms. OpenMP can be used instead of MPI. The partitioning allows the analysis at all levels given the communications level, speed, and protocol at that level. Consequently, the present solution is suitable for high-performance computers with shared or distributed memory, clusters of computing nodes with high speed networks and potentially P2P networks with heterogeneous nodes. One technique to further improve performance is to reduce the search space by choosing partitions where an optimal alignment is likely to be found. Further optimal or near optimal solutions can be retrieved by forking at different equally ranked paths, or within a threshold defined by the user. Other high dimensional scientific computing problems can be approached similarly, with the goal of producing a reusable mathematical model to represent partitioning and dependency in parallel algorithms generally, and high dimensional problems specifically.

## 7. Acknowledgments

The authors wish to thank Vitali Sintchenko for helpful comments. Funding from The University of Sydney for Sun Fire X2200 M2 Server with 2xAMD Opteron quad core is acknowledged.

## 8. References

[1] Wong, K. M, et al., Alignment uncertainty and genomic analysis. *Science*. 319(5862), pp. 473-6, 2008.
[2] Golubchik, T., et al .,Mind the gaps, evidence of bias in estimates of multiple sequence alignments. *Mol. Biol. and Evol*. 24(11), pp. 2433-42, 2007.
[3] Giegerich, R., Meyer, C., Steffen, P.,A discipline of dynamic programming over sequence data. *Science of Computer Programming*. 51, pp 215-63, 2004.
[4] Raynolds, J., Mullin, L. M., Applications of conformal computing Techniques to Problems in computational physics, the FFT. *Comp. Phys. Comm*. 170, pp. 1-10, 2005.
[5] Helal, M., Mullin, L. M., Gaeta, B., El-Gindy, H. , Multiple sequence alignment using massively parallel mathematics of arrays. *In, Proceedings of the International Conference on High Performance Computing, Networking and Communication Systems (HPCNCS- 07)*. Orlando. FL. USA, pp. 120-7,2007.
[6] Stone, H., Multiprocessor scheduling with the aid of network flow diagram. In, *IEEE trans. on Soft. Eng*. SE-3. pp. 85-93,1977.
[7] Thompson, J. D., Higgins, D. G., Gibson, T. J. , CLUSTAL W, improving the sensitivity of progressive multiple sequence alignment through sequence weighting, position specific gap penalties and weight matrix choice. *(http://wwwbimas.cit.nih.gov/clustalw/clustalw.html)*.
[8] Notredame, C., Higgins, D. G., Heringa1, J. , T-Coffee: A Novel Method for Fast and Accurate Multiple Sequence Alignment. *J. Mol. Biol*. pp. 205-17,2000.
[9] Edgar, R. C.,MUSCLE: multiple sequence alignment with high accuracy and high throughput. *Nucleic Acids Research*. 32(5), pp.1792-7,2004.